\documentclass[letterpaper]{article}
\usepackage[]{aaai24}
\usepackage{times}  
\usepackage{helvet}  
\usepackage{courier}  
\usepackage[hyphens]{url}  
\usepackage{graphicx} 
\usepackage{natbib}  
\usepackage{caption} 
\usepackage{subcaption}
\usepackage{booktabs}
\usepackage{multirow}
\usepackage{amsmath}
\usepackage{url}
\usepackage{float}

\usepackage{amssymb}
\usepackage{tabularx}
\usepackage{makecell}
\usepackage{multicol}

\usepackage[T1]{fontenc}
\usepackage[utf8]{inputenc}

\usepackage{microtype}

\usepackage{inconsolata}

\newcommand{\figref}[1]{Figure \ref{#1}}

\newcommand{\tabref}[1]{Table \ref{#1}}
\newcommand{\secref}[1]{Section \ref{#1}}

\usepackage{color}

\usepackage{newfloat}
\usepackage{listings}
\DeclareCaptionStyle{ruled}{labelfont=normalfont,labelsep=colon,strut=off} 
\floatstyle{ruled}
\newfloat{listing}{tb}{lst}{}
\floatname{listing}{Listing}
\title{Does My Dog ``Speak'' Like Me? 
The Acoustic Correlation between Pet Dogs and Their Human Owners}
\author {
    Jieyi Huang\textsuperscript{\rm 1},
    Chunhao Zhang\textsuperscript{\rm 2},
    Yufei Wang\textsuperscript{\rm 3},
    Mengyue Wu\textsuperscript{\rm 4},
    Kenny Zhu\textsuperscript{\rm 5*}
}
\affiliations {
    \textsuperscript{\rm 1,2,3,4}Shanghai Jiao Tong Univerisity\\
    \textsuperscript{\rm 5}The University of Texas at Arlington\\
    \textsuperscript{\rm 1}xsiling1@gmail.com, \textsuperscript{\rm 2}forest\_zch@sjtu.edu.cn, \textsuperscript{\rm 3}arthur-w@sjtu.edu.cn\\
    \textsuperscript{\rm 4}wumengyue@cs.sjtu.edu.cn, \textsuperscript{\rm 5}kenny.zhu@uta.edu
}

\begin{document}
\maketitle

\begin{abstract}
How hosts language influence their pets' vocalization is an interesting yet underexplored problem.
This paper presents a preliminary investigation into the possible correlation 
between domestic dog vocal expressions and their human host's
language environment.
We first present a new dataset of Shiba Inu dog vocals from YouTube, which 
provides 7500 clean sound clips, including their contextual information
of these vocals and their owner's speech clips with a carefully-designed data processing
pipeline. The contextual information includes the scene category 
in which the vocal was recorded, the dog's location and activity. 
With a classification task and prominent factor analysis, 
we discover significant acoustic differences in the dog vocals from the 
two language environments. We further identify some acoustic features from 
dog vocalizations that are potentially correlated to their host language patterns\footnote{The data, code
	and a live demo are available at 
	\url{https://anonymous.4open.science/r/EJSHIBAVOICE-0DC5}.}.
\end{abstract}

\section{Introduction}
\label{sec:intro}

Understanding animals' verbal expressions is an interesting interdisciplinary 
scientific challenge. This is  particular true with pet dogs, 
who closely interact with humans. 
Previous research endeavors to comprehend dog vocal sounds for a number 
of reasons, such as for a better understanding of animal biological 
evolution~\cite{pongracz2017modeling}, applying their language to information 
technology, or just curiosity about dogs' intention when they make a sound~\cite{pongracz2011children,dogbark_1}. However, this task is challenging not only due to the unknown acoustic pattern of dogs but also the lack of a suitable and high-quality dataset.

\begin{figure}[th]
	\centering
	\scalebox{0.24}{\includegraphics{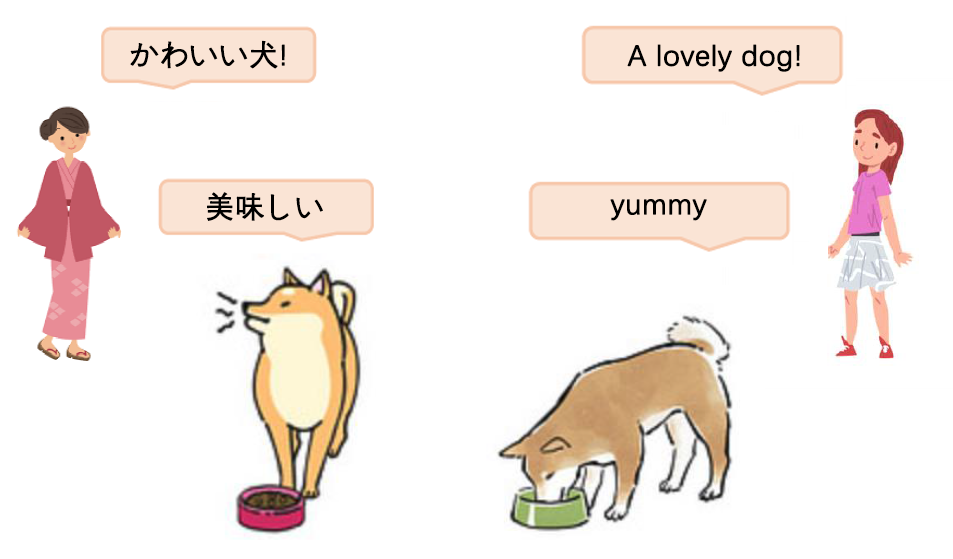}}
	\caption{
A dog's vocalization may be accousticly correlated with its host language, 
so that a ``Japanese'' dog should vocalize differently than an ``American''
dog under the same context such as ``eating on the lawn''.}
\label{fig:intropic}
\end{figure}

Previous researchers have demonstrated that a dog's vocalization indeed reflects 
their individual characteristics~\cite{pongracz2010barking,larranaga2015comparing}, emotional expression~\cite{thorndike2017animal,hantke2018my,paladini2020bark} and perception of outside world~\cite{larranaga2015comparing, molnar2008classification}. However, despite the fact that dogs are human's most familiar animals, 
little research has looked into the influence on dogs' communication arising from their interaction with human hosts. As a matter of fact, dogs exhibit many modes of 
communication ranging from behavioural patterns to vocalizations, in this paper we pay attention to their vocal sounds, which serve as one of 
the most important communication channels~\cite{siniscalchi2018communication}. 
In our work, we hypothesize that a dog's acoustic characteristics may be 
correlated with such interaction, particularly the host's spoken language. 
To verify that, we explore the vocal difference of a particular dog breed 
(Shiba Inu) from two different host language environments (English vs. Japanese)~(\figref{fig:intropic}). Shiba Inu dogs are chosen as the subject of this study because Shiba Inus are very popular among dog owners and there is an abundance of their audio/video resources available online.  Moreover, we choose to work on dogs in English and Japanese language environments because these two widely-spoken languages have very different phonetic systems, and at the same time, Shiba Inu dogs are very popular in both Japanese and English-speaking households (i.e., in Japan and in the US). 

One may argue that the host language is not the 
only factor that might influence how dogs sound in the online videos. 
For example, social norms and customs play also play apart in these 
countries/cultures. We believe these factors are inter-related with the 
languages that are in question in this paper. 
Having correlation between dog vocals and human speeches does not nullify the correlation between dog vocals and social norms and cultural behaviors, and vice versa. In fact, these two kinds of correlations can strengthen each other.

We verify the above hypothesis via a two-stage pipeline. 
First, we conduct classification experiments to investigate the possible
existence of interesting acoustic properties that distinguish dog vocalizations from one language environment to the other. The classification experiment is 
performed on pairs of Japanese and English dog vocal clips under the same \textit{context} which is composed of the \textit{scene category}, \textit{location}, and \textit{activity} of the dog during the recording, so as to exclude  
these confounding non-linguistic factors, which may affect how dogs voice out. 

In stage two, to discover the most prominent factors that distinguish dog vocals 
by their language environments, we perform an importance analysis of different 
factors using Shapley values. A similar analysis is also performed on host 
languages, to study the acoustic difference between English and Japanese. 
Moreover, we compute the Pearson correlation between dog vocals and their host 
speech.  The fact that several most important acoustic characteristics differentiating dogs
have substantial correlations with those of their host language 
(i.e., English or Japanese) supports our hypothesis that 
the domestic dog vocal expressions do share a few acoustic similarities 
with its host language. It is possible that the host language environment has 
an influence over the dog's vocalization. 

Previous dog vocalization datasets are mostly collected under controlled evironment, i.e. the researchers raised several dogs and recorded their physical  as well as vocal behaviours~\cite{ide2021rescue, ehsani2018let, 
molnar2008classification, hantke2018my}. However, such data acquisition methods are not only costly but also lacking generalization capability as the number of dogs is highly restricted. 
Since webly data provides a large amount of dog vocalizations with plenty of metadata featuring the dog breed, language environment and contextual information, we derive a pipeline to obtain and filter the vocals from social media, leading to a large-scale of dog vocalization dataset ``EJShibaVoice''.
The number of families for Japanese and English respectively stand at 219 and 275, much larger than any other similar studies previously reported. 
Moreover, we use the context information tagged with the vocal clips to eliminate the confounding factors in our experiments. 
Specifically, we develop 
a framework that crawls Shiba Inu audio clips from both English and 
Japanese-speaking host families, extracts the vocal clips, 
segments the clips into contiguous singular sounds, and tags them with 
contextual information.

Our main contributions are summarized as follows:

\begin{itemize}
	\item A newly-defined task to uncover the human linguistic influence 
on the vocal expressions of domestic Shiba Inu dogs via a unique data-driven and computational approach, 
which can inspire further research on animal 
languages.~(\secref{sec:assumption}) 
	\item We construct a large-scale Shiba Inu vocalization dataset \textbf{EJShibaVoice} containing clean audio clips produced by dogs from two different language environments: English and Japanese, including dogs' host speech clips. 
The dog vocal and human speech data undergo a systematic pipeline, 
which extracts clean dog voices and the corresponding host speech from 
the social media videos.~(\secref{sec:assumption})
	\item We discover prominent acoustic differences between dogs 
from different language environments: 
	Shiba Inus vocalizations from English-speaking households have a \textbf{\textit{lower frequency}}, 
while those from Japanese environments have \textbf{\textit{faster speed}}, 
which correlates with these two human languages, respectively.~(\secref{sec:main})
\end{itemize}

\section{Problem and Dataset}
\label{sec:assumption}
In this paper, we ask two research questions: 
\begin{enumerate}
\item Do pet dogs from different human language environments sound differently? 
\item If so, is their vocalization related to their host's language in any way? 
\end{enumerate}

To answer these questions, we construct a dataset ``EJShibaVoice'', 
which is composed of Shiba Inu vocal samples from Japanese and English language environments 
and their corresponding hosts speech. Additionally, each audio sample is also tagged with
metadata such as the begin and end timestamps, and the context under which the vocal sound is made.    
Additionally, host's speech from the same video is extracted using a similar approach as we process dog vocalizations. This procedure will be described in detail in~\secref{sec:speechextraction}.

To ensure the high quality of the dataset, we develop a rigorous processing pipeline to extract pure dog sounds in a variety of environments, with little or no extraneous noises such as human speech or background music. We segment dog audio clips into minimal units, which is a full singular bark, for the purpose of fine-grained acoustic comparison between different language environments. 
We further provide detailed extraneous context for each vocal clip, as there are other variables other than language environments that could potentially influence dog's  sound, for example whether a dog is eating or playing, outdoor or indoor\cite{larranaga2015comparing, molnar2008classification}. We make this fine distinction on the context of the dog sounds so as to constrain confounding factors other than the linguistic environment of the dogs. Hereby, with the sound segmentation and context tagging, each clean singular vocalization has a corresponding context.

\begin{figure}[ht]
	\centering
\begin{subfigure}[t]{0.48\columnwidth}
        \centering
        \includegraphics[width=0.98\columnwidth]{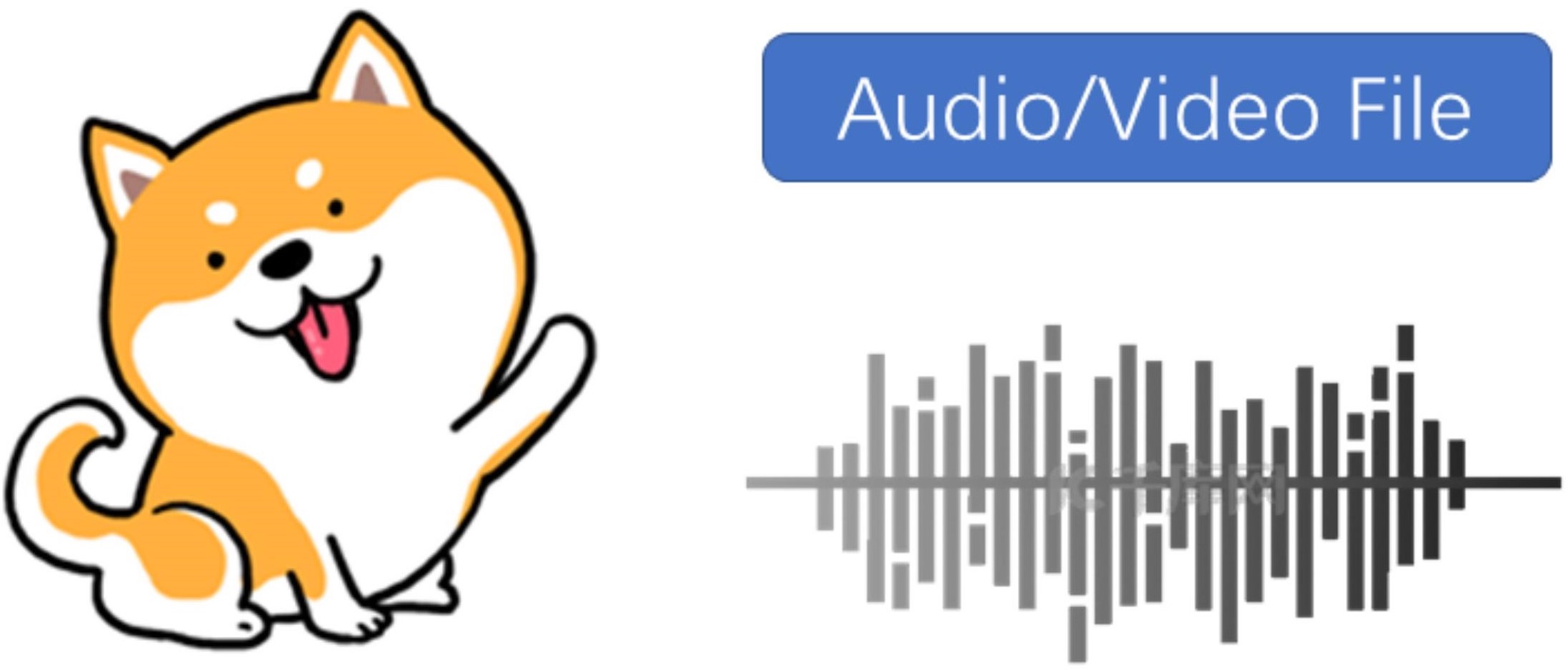}
        \caption{source videos and tag language.}
        \label{fig:source}
\end{subfigure}
\begin{subfigure}[t]{0.48\columnwidth}
        \centering
        \includegraphics[width=0.7\columnwidth]{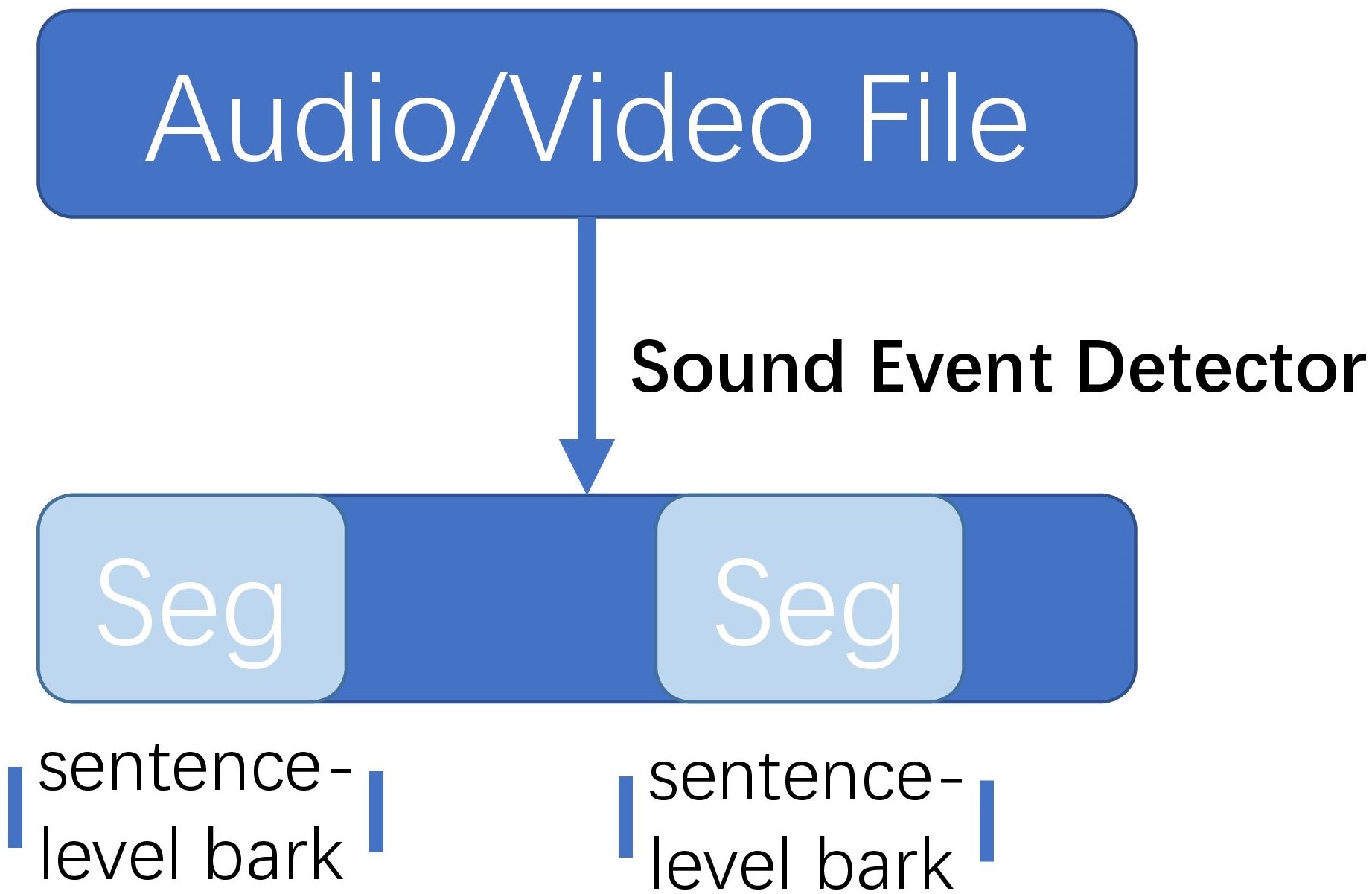}
	\caption{sentence-level segmentations.}
        \label{fig:sentence}
\end{subfigure}
\begin{subfigure}[t]{0.48\columnwidth}
        \centering
        \includegraphics[width=0.7\columnwidth]{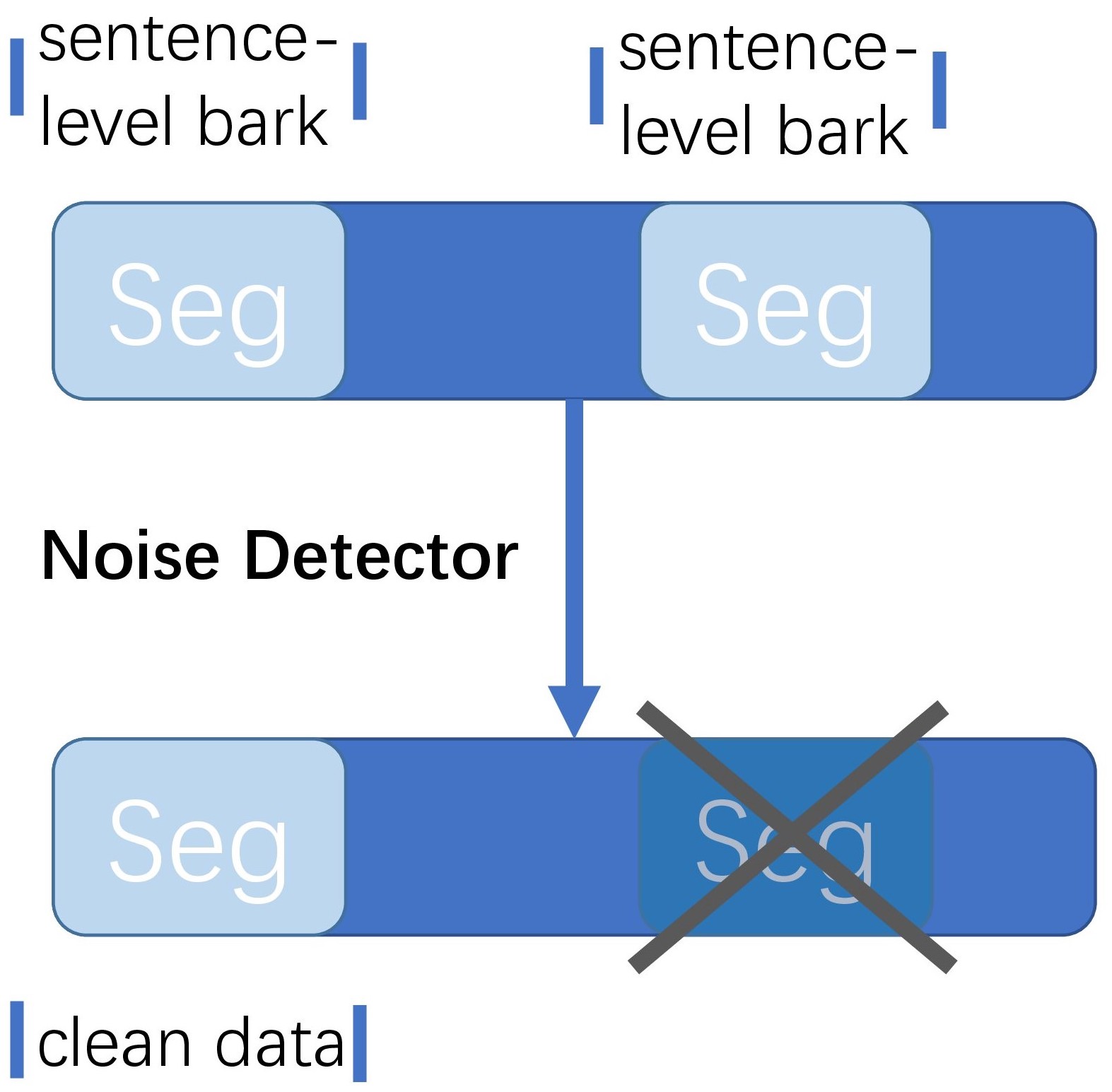}
	\caption{clean data after noise removing.}
        \label{fig:clean}
\end{subfigure}
\begin{subfigure}[t]{0.48\columnwidth}
        \centering
        \includegraphics[width=0.98\columnwidth]{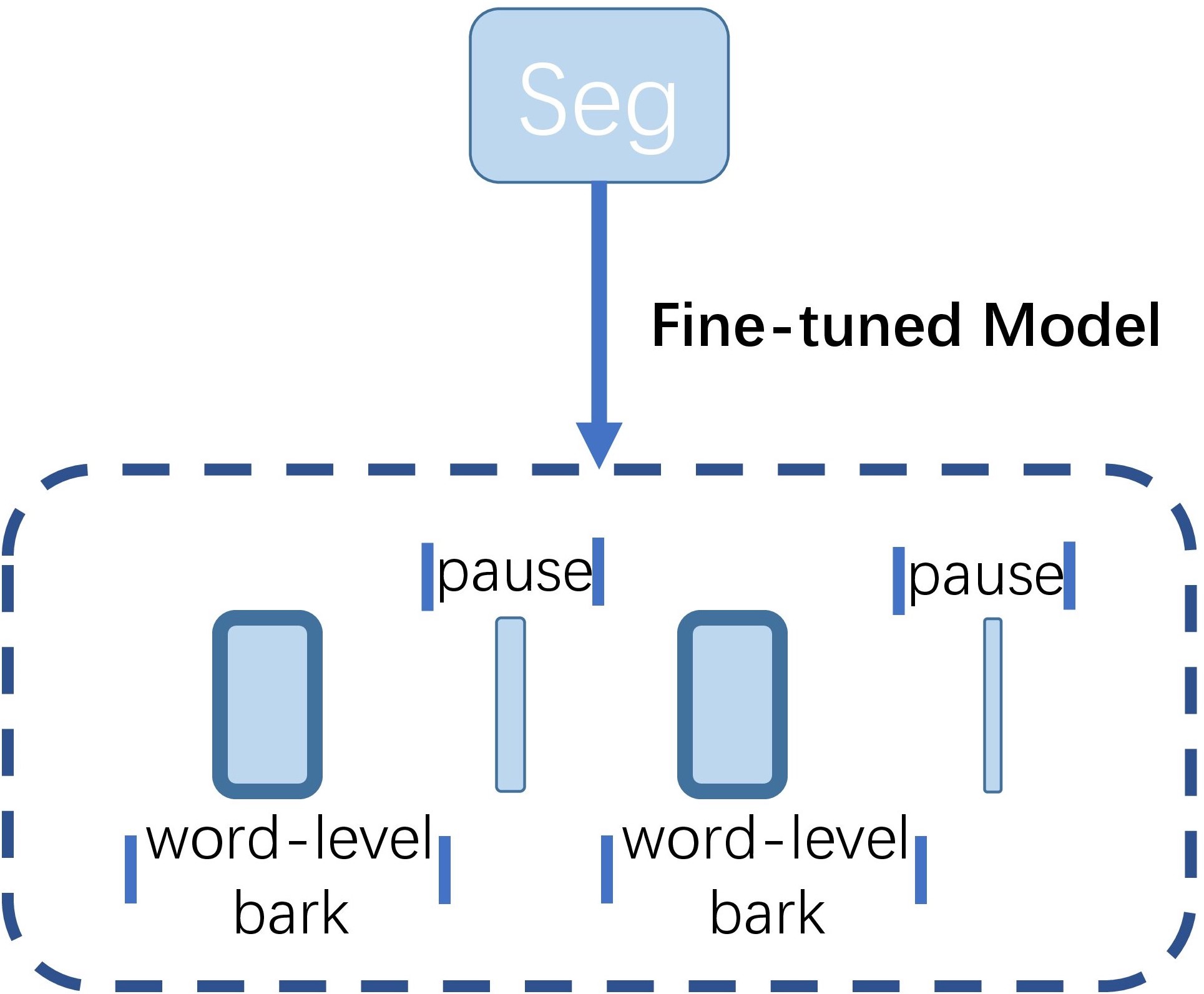}
	\caption{word-level barks.}
        \label{fig:word}
\end{subfigure}
\begin{subfigure}[t]{0.7\columnwidth}
        \centering
        \includegraphics[width=0.98\columnwidth]{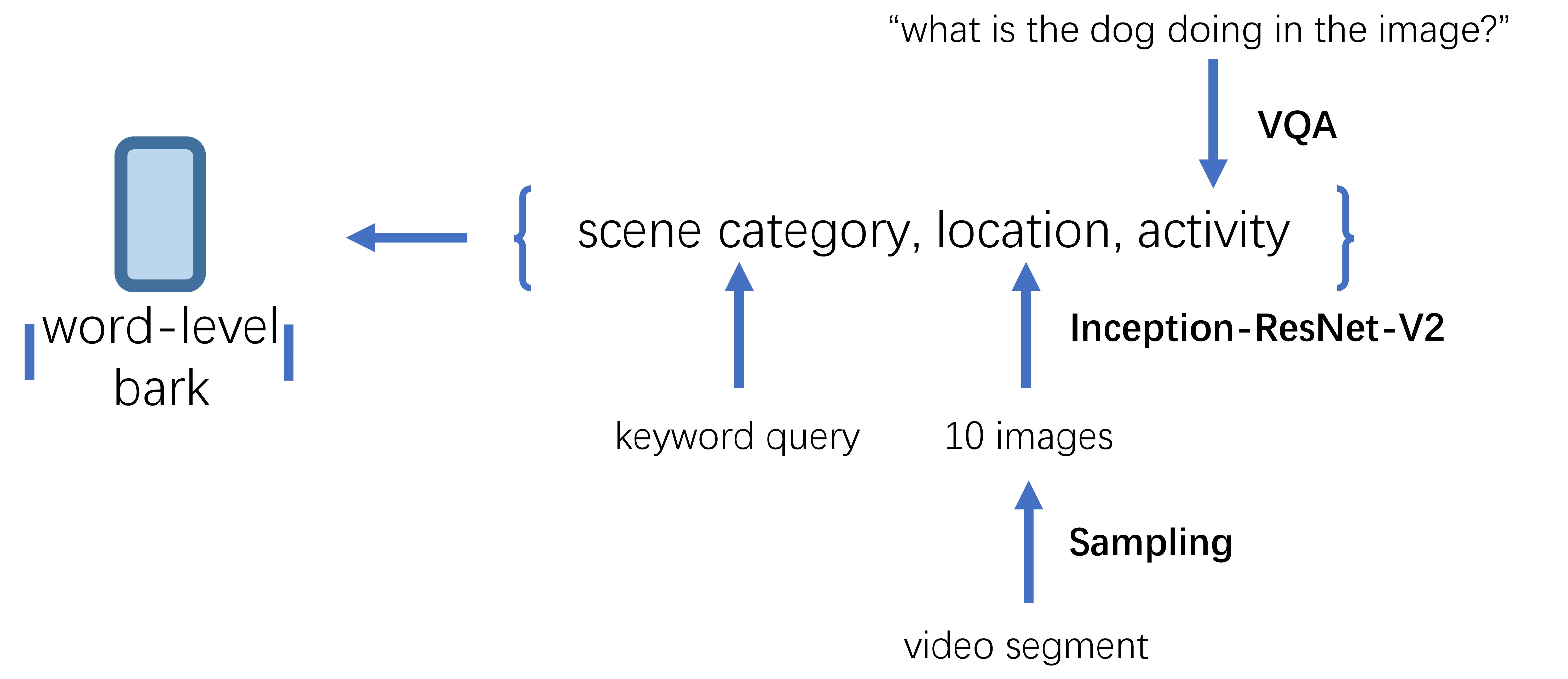}
	\caption{context tagging.}
        \label{fig:context}
\end{subfigure}
\caption{The procedure of dataset construction.}
\label{fig:methodpic}
\end{figure}

Next, we present the full data processing pipeline~(\figref{fig:methodpic}) 
for constructing the EJShibaVoice dataset. 

\subsection{Sourcing Dog Vocalization Sounds Online}
\label{sec:source}
Since YouTube contains a large amount of user-uploaded videos about their pet dogs, our first step is to source relevant clips from two language environments via searching for the keyword
``Shiba Inu'' and a scene category term on YouTube. In this study, we use a total of eight scene
categories, which are defined in \tabref{tab:context}. These eight categories were proposed
in previous work about dog activity classification~\cite{larranaga2015comparing}. 
For example, to download videos about Shiba Inu eating, we search for
the query ``Shiba Inu Eat -Coin'' for English and the correponding Japanese of this query for vocals under Japanese host language environment. Here ``-Coin'' is used as there are a large number of Shiba Inu Coin videos, we add this term to filter out those irrelevant videos. All eight scenes are searched under the same keyword patterns.

Because YouTube videos are not typically tagged with original language or locale, we infer the language environment of a video by the languages ofits caption or title. For example, if the title contains Japanese characters, we consider it as a video recorded in an Japanese-speaking environment, whereas a video with purely English title is considered to be recorded in an English-speaking environment. 

After all the videos are downloaded and tagged, 16,161 videos in total are collected. Among them we randomly sample 1,000 clips to verify the host language tagging accuracy by watching the video clips. The accuracy is 93.2\%, which indicates that the majority of the Shiba Inu videos are tagged with the correct host language environments, and fit our research purpose.

\subsection{Dog Vocalizations Extraction}
\label{sec:barkextraction}
The original audios downloaded from YouTube contain long uninformative segments where dogs are silent or background noises muffling the vocalizations. To ensure the quality of 
our dataset, we have adopted a systematic and rigorous pipeline of three steps 
to extract pure and singular vocal segments from the audios.

\paragraph{Step 1: sentence-level segmentations extraction}
As a preliminary step, we extract long segments in which contains a continuous series of
dog vocalizations. Such long segments, which we call a dog ``sentence'', can be detected because
they are preceded or followed by significant periods of silence. 
We apply PANNs~\cite{kong2020panns}, a pre-trained large sound 
event detection model including as many as 527 sound classes for recognizing the 
sound events. The continous segments detected by PANNs with the event class ``barking'' are considered as the sentence-level bark segmentations.
\paragraph{Step 2: noise elimination}
From our practical experience, we have found that sometimes vocalizations will be unclear 
due to the background music and human talking. Thus in the second step, 
among those sentence-level segmentations, those with co-existing event ``speech'' and 
``music'' detected by PANNs are removed from the dataset.

\paragraph{Step 3: word-level vocalizations extraction}
Another challenge is that the coarse-grained sound clips may have 
some short pauses in the middle. To remove 
these pauses, we finetune a sound event detection model to 
determine the start and end time of singular vocalizations, which we call ``words'' 
in this paper.

PANNs is used as the pretrained model, which is trained on AudioSet~\cite{gemmeke2017audio} for all the sound classes first. 
We then manually made framewise labels for the event ``barking'' on 246 audio clips with a total length of 715 seconds and fine-tuned this pretrained model.
This fine-tuned model can precisely detect dog vocals in the audio with the start time and end time. Based on the precise start time and end time, we use the fine-tuned model to remove the pause segments between barkings.  

At the end of these three steps, only a singular dog vocalization is left in each clip. 
In total, 7,500 clips from 1,551 raw videos remained. While we acknowledge the fact that the data noise induced by recording devices and audio ambience is difficult to eliminate via web data, such factors/biases should be well mitigated with the large and diverse data source coming from either side of the Pacific that we curated on YouTube.

\subsection{Context Metadata Tagging}
\label{sec:tag}

We list the range of possible values for {\em scene}, {\em location}
and {\em activity}, three components of the context in \tabref{tab:context}. The set of values for locations values is a subset of the location set from Inception-ResNet-V2 model. 

\begin{table}[th]
\small
\begin{tabular}{p{0.12\columnwidth}|p{0.8\columnwidth}}
\toprule
\textbf{Context} & \textbf{Possible values} \\ \midrule
Scene & Alone, Bath, Eat, Fight, Play, Run, Stranger, Walk \\ \midrule

Location & 76 possible locations, such as lawn, street, office, recreation\_room and so on \\ \midrule
Activity & a 768-dimentional real valued vector \\ 
\bottomrule
\end{tabular}
\caption{Description of the context metadata. 
}
\label{tab:context}
\end{table}

\begin{figure}[th]
	\centering
	\includegraphics[width=\columnwidth]{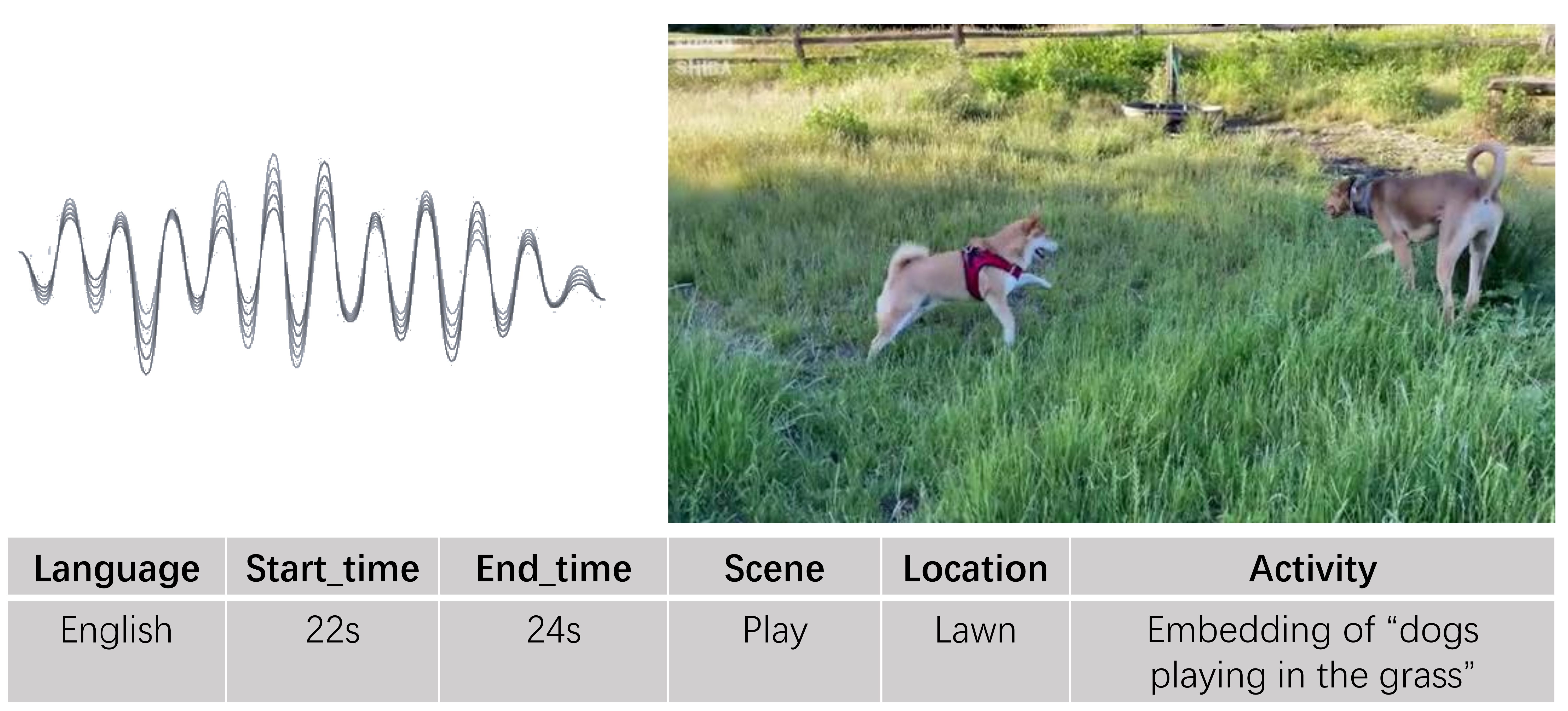}
	\caption{One audio sample in EJShibaVoice with its metadata of language environment, 
timestamps, and context.} 
	\label{fig:EJShiba}
\end{figure}

For locations and activites, we sample 10 images (equi-spaced) from the video segment that 
coincides with the duration of the audio bark sample, and classify the images into one of the location labels listed in \tabref{tab:context}. The locations are inferred from Inception-ResNet-V2 model\cite{szegedy2017inception} trained on AI Challenge 2017 Scene Classification dataset. We apply image caption and visual question answering (VQA) models from OFA~\cite{wang2022unifying} to first generate a caption for an image and then ask the model ``what is the dog doing in the image?''. The caption results are transformed into word embeddings with pre-trained BERT model\cite{devlin2018bert}, and the 10 word embeddings of the 10 images are averaged to obtain the overall activity embedding of that bark sample. Besides, the timestamp of each bark clip in the original video is provided. Some samples from this dataset are shown in \figref{fig:EJShiba}.

\subsection{Host Speech Extraction}
\label{sec:speechextraction}
To address the second research question, which is to find
the correlation between the barks and their host languages, 
we need the audio speech data from dogs' hosts as well. 
The human speech in the original audio track is also extracted by a 
strict pipeline. The first few steps are similar to dog barking extraction: 
we start by getting those continuous segments tagged as ``speech'' by PANNs and 
remove those tagged as ``music'' from them to reduce the noises. As it is normal for several people talking simultaneously, we further apply a speaker diarization 
model Pyannote\cite{Bredin2020, Bredin2021} to separate them from each other and retain speech of a single person 
in more fine-grained segments.

\subsection{The Final Product}
The dataset consists of two parts: the barks and the corresponding host speech. After the whole processing, 7,500 clean bark sounds and 15,197 corresponding host speech clips 
finally remain in our EJShibaVoice dataset~(\tabref{table:datasetlength}).

\begin{table}[th]
	\tiny
	\centering
	\begin{tabular}{c|c|c|c|c}
		\toprule
		{}            & \# of Clips & Avg Len (s) & Var of Len ($s^2$) & English Per(\%)\\
		\midrule
		{Bark} & 7500 & 0.61  &  0.288 & 46.72\\
		{Speech} & 15197 & 1.56 & 2.134 & 44.01\\
		\bottomrule
	\end{tabular}
	\caption{EJShibaVoice Statistics.
}
	\label{table:datasetlength}
\end{table}

As the data statistics indicate, we include roughly equal number of audio clips 
from English and Japanese environments, leading to a balanced dataset that can well 
support our investigation on the influence of language environment.

\begin{figure}[th]
	\centering
	\scalebox{0.2}{\includegraphics{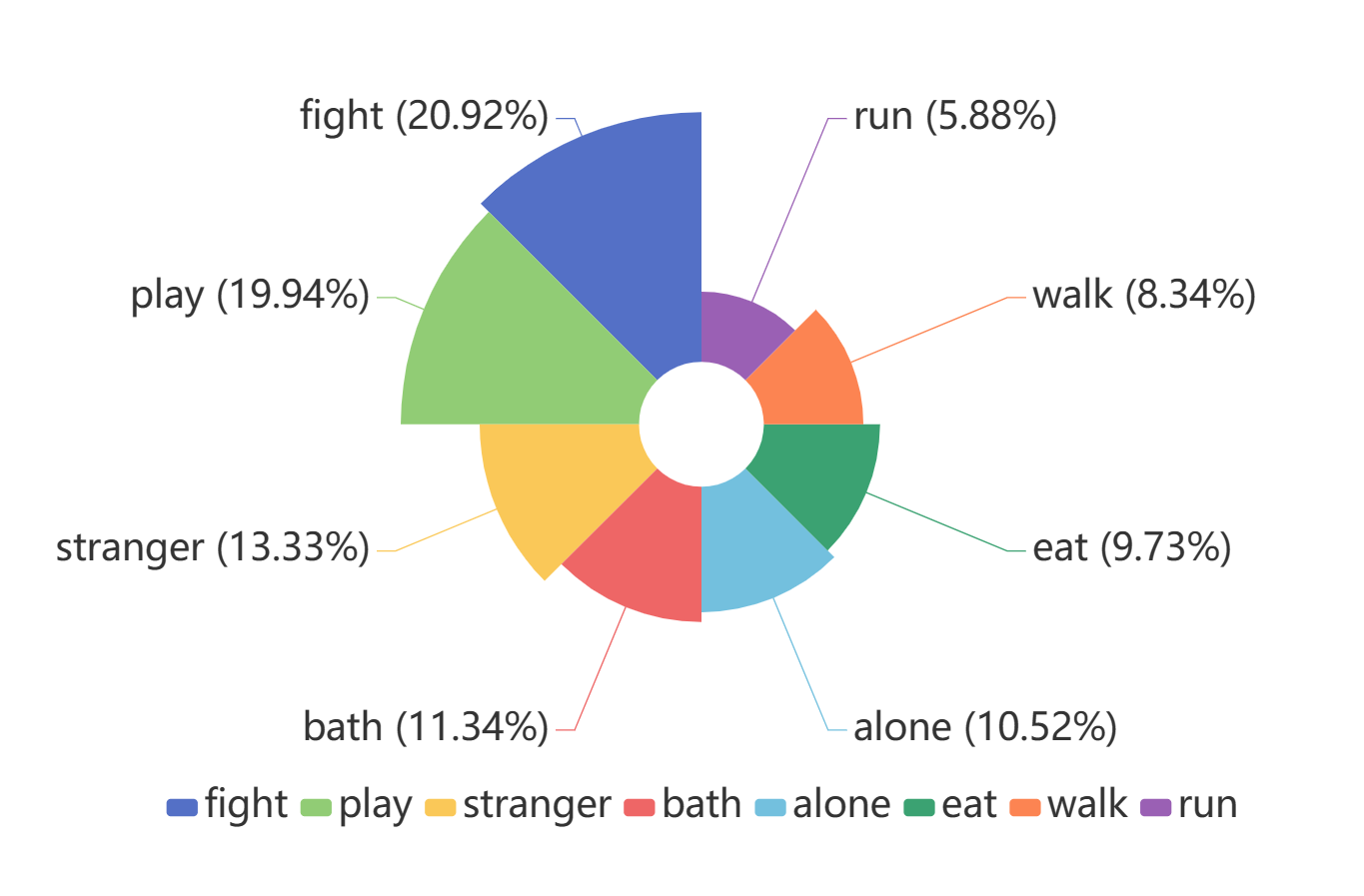}}
	\caption{The percentage of bark clips for different scenes.}
	\label{fig:keyword_rosepie}
\end{figure}

The number of clips of each scene differs~(\figref{fig:keyword_rosepie}). 
As we use the same method to process each scene, the possible reason for imparity is that Shiba Inu barks diversely under different conditions. For example, dogs tend to bark more
to show their strength when fighting, therefore the number of clips under ``fight'' is 
the largest.

\section{Methodology}
\label{sec:method}
To verify our assumptions that dogs from different human language environments have different vocals and the difference is related to their host's language, we conduct classification-based experiments and analyze the Shapley value to find which features are important in distinguishing their vocals. 

\subsection{To Answer the First Question: Pairwise Classification}

To better control context and language environment, we adopt a 4-way classification for pairwise comparison. Specifically, we pair the dog vocalizations from the same context but under different language environments in 4 ways: En-En, Ja-Ja, En-Ja, and Ja-En. Two clips are considered to be from the same context if they have the same scene category, the same location, and their activity vectors with a cosine similarity of 0.95 or above. The problem is then defined as the classification of these pairs into any of the above four classes. Our classification models include xgboost, KNN, Logistic Regression, and Random Forest. Multiple acoustic features including spectral features (filterbank~\cite{strang1996wavelets}, PLP~\cite{hermansky1990perceptual}, MFCC~\cite{davis1980comparison}) and handcrafted feature-sets (eGeMAPs, GeMAPs~\cite{eyben2015geneva}, ComParE~\cite{schuller2013interspeech}) are utilized to feed into the model. 

Dogs may also differ by their
age and sex, but these attributes are hard to obtain from the YouTube data even manually.
Given the large size of our dataset, we believe a significantly higher than random classification
accuracy will show that we can distinguish English dogs with Japanese dogs just 
by acoustic features.  

An alternative method is to group all fine-grained  clips by the contexts, and then
do a two-way (English or Japanese environment) classification in each group using only 
acoustic features.  However given the complex combination of  scene, location and activity, few samples from both languages share exactly the same context hence we conducted the more difficult four-class experiment. 

\subsection{To Answer the Second Question: Correlation on Prominent Factors}
To ascertain the influence of the host language on dog vocalizations, we analyze prominent factors 
that distinguish Japanese and English dogs' sounds. Shapley value is commonly adopted 
to explain feature importance for a given machine learning model, which can help 
determine the prominent features influencing dog vocalizations. 



To compare the relationships between a dog vocalization and the host language, 
we also include features extracted from human speech (English and Japanese corpus). 
The speech sources include 8,000 clips from CommonVoice~\cite{ardila2019common}, 
which is an open source multilingual speech dataset contributed by volunteers around 
the world, and host speech from EJShibaVoice. We adopt two different sets of speech data 
because they provide distinctive features. 
Speech from EJShibaVoice has direct relation with vocalizations, 
so that we can conduct Pearson value analysis between them, 
while CommonVoice is purer and more common to help us find universal 
feature of human speech. Similar procedures are conducted on human language 
and the prominent factors are later compared with those inferred 
from dog vocalizations~(\secref{sec:prominentfactor}).
Furthermore, to ascertain the correlation between vocalizations and their host speech in a statistical way, we analyze the Pearson correlation between them. In the meantime, the Pearson correlation between vocalizations and random speech is shown to compare~(\secref{sec:prominentfactor1}).

\section{Results and Analysis}
\label{sec:results}

In this section, we present the classification results from different machine learning models with different features extracted from our dataset EJShibaVoice. We further compare Shapley values on the GeMAPs feature to find the prominent factors. We also try several different classification models on both spectral and handcrafted acoustic features.

\subsection{Results of Pairwise Classification}
\label{sec:main}
\begin{table}[th]
	\small
	\centering
	\begin{tabular}{l|l|c|c|c|c}
		\toprule
		\multicolumn{2}{c|}{}            & xgboost & KNN  & LR   & RF  \\
		\midrule
		\multicolumn{2}{c|}{filterbank}          & 0.9827        & 0.9057         & 0.6374    & 0.9603    \\
		\multicolumn{2}{c|}{PLP} & 0.9733 & 0.7701 & 0.4375 & 0.9123\\
		\multicolumn{2}{c|}{MFCC} & 0.9828 & 0.9161 & 0.5441 & 0.9587\\
		\multicolumn{2}{c|}{ComParE} & {0.9868} & {0.5717} & {0.6004} & {0.9520}\\
		\multicolumn{2}{c|}{GeMAPs} & 0.9836 & 0.7317 & 0.6230 & 0.9474 \\
		\multicolumn{2}{c|}{eGeMAPs} & 0.9840 & 0.7432 & 0.6901 & 0.9567\\

		\bottomrule
	\end{tabular}
	\caption{4-class classification accuracy on the vocal pairs.}
	\label{table:mainresult}
\end{table}


In total, we form 9,200 pairs of vocal clips. Among these, we formed 4 different pairs, EN-EN, JA-JA, EN-JA, and JA-EN, with 2,300 clips in each set respectively.  
And we perform 5-fold cross-validation on this paired dataset. 
The overall classification results are presented in \tabref{table:mainresult}, 
where six commonly-adopted audio features are compared using four different classification models. Among these features, filterbank, PLP, and MFCC 
are extracted from the spectral transformation and have 24, 13, and 13 dimensions respectively. By contrast, ComParE, GeMAPs, and eGeMAPs are human-crafted features and have 6373, 62, and 88 dimensions respectively. 
These features are easier to explain from perception perspectives, 
however, they might be less informative than direct spectral features, 
resulting in relatively lower accuracy in some models.

Our first observation is that no matter which feature set or which model is used, 
dog vocals from different host language environments can be clearly distinguished, 
because the accuracies are all clearly higher than 0.25 (the random guess).
In other words, this indicates that there is a certain acoustic difference 
between the dog vocals in these two language environments. 
Specifically, the accuracy of ComParE significantly drops while using KNN, 
which is due to the curse of dimensionality of this 6,373 dimension feature.
XGBoost shows the highest classification accuracy, 
with all six features demonstrating an accuracy higher than 0.90. This result affirms that
\textbf{dog vocals under English environments are distinctly different from 
those under Japanese environments}. 

Next, we will find out what are the crucial features 
that distinguish the English dogs from the Japanese dogs. 


\subsection{Results of Correlation on Prominent Factors}
\subsubsection{Prominent Analysis on GeMAPs}
\label{sec:prominentfactor}

\begin{figure}[th]
	\centering
	\scalebox{0.4}{\includegraphics{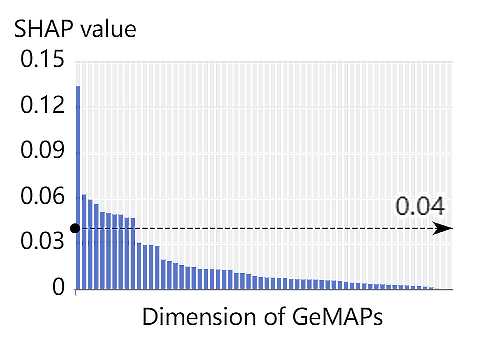}}
	\caption{The average SHAP values (absolute value) of GeMAPs sorting from high to low. Features with SHAP values higher than 0.04 are considered prominent because of the 
sudden drop in SHAP values to the right those features. 
}
	\label{fig:prominentfig} 
\end{figure}

The results of SHAP values can be seen in \figref{fig:prominentfig} and details of the selected dimensions are listed in \tabref{table:prominentfactor}.

\begin{table}[th]
	\scriptsize
	\centering
	\begin{tabular}{l|c|c}
		\toprule
		Dimension & Dim Type & SHAP  \\
		\midrule
		loudness\_sma3\_amean & Energy &  0.1336     \\
		F0semitoneFrom27.5Hz\_sma3nz\_percentile50.0 & Frequency &  0.0622  \\
		loudness\_sma3\_meanRisingSlope & Energy &  0.0590  \\
		logRelF0-H1-A3\_sma3nz\_stddevNorm  & Frequency & 0.0561  \\
		loudnessPeaksPerSec& Temporal  & 0.0507 \\
		F0semitoneFrom27.5Hz\_sma3nz\_percentile80.0 & Frequency  & 0.0498  \\
		hammarbergIndexV\_sma3nz\_stddevNorm & Spectral & 0.0491 \\
		slopeV0-500\_sma3nz\_amean & Temporal & 0.0490 \\
		loudness\_sma3\_percentile80.0& Energy & 0.0470  \\
		slopeV500-1500\_sma3nz\_stddevNorm& Temporal  & 0.0467\\
		\bottomrule
	\end{tabular}
	\caption{Details of Prominent Dimensions in GeMAPs. }
	\label{table:prominentfactor}
\end{table}
The ten prominent factors in \tabref{table:prominentfactor} fall into four categories: 
\textit{spectral}, \textit{temporal}, \textit{energy}, and \textit{frequency}, 
according to the original GeMAPS definition. Most of the prominent factors are
spectral parameters, including dimensions of HammerbergIndex and slope. 
HammerbergIndex represents the ratio of the strongest peak in the 0-2kHz region to 
the strongest peak in the 2-5kHz region, while Slope represents the linear regression slope 
of the spectral power spectrum within the given band. F0semitone-related factors describe the pitch, which is highly related to the
fundamental frequency. The factors about segment length are temporal parameters. 
The energy-related parameter loudness, which represents the sound intensity, 
is largely affected by the recording device and environment. 


Considering these factors, the results show that dog vocals from two host language environments have distinctive differences in their energy distribution over frequency. In quantitative analysis, vocals from the Japanese language environment have a higher frequency than those from English.

\begin{figure}[th]
	\centering
	\scalebox{0.24}{\includegraphics{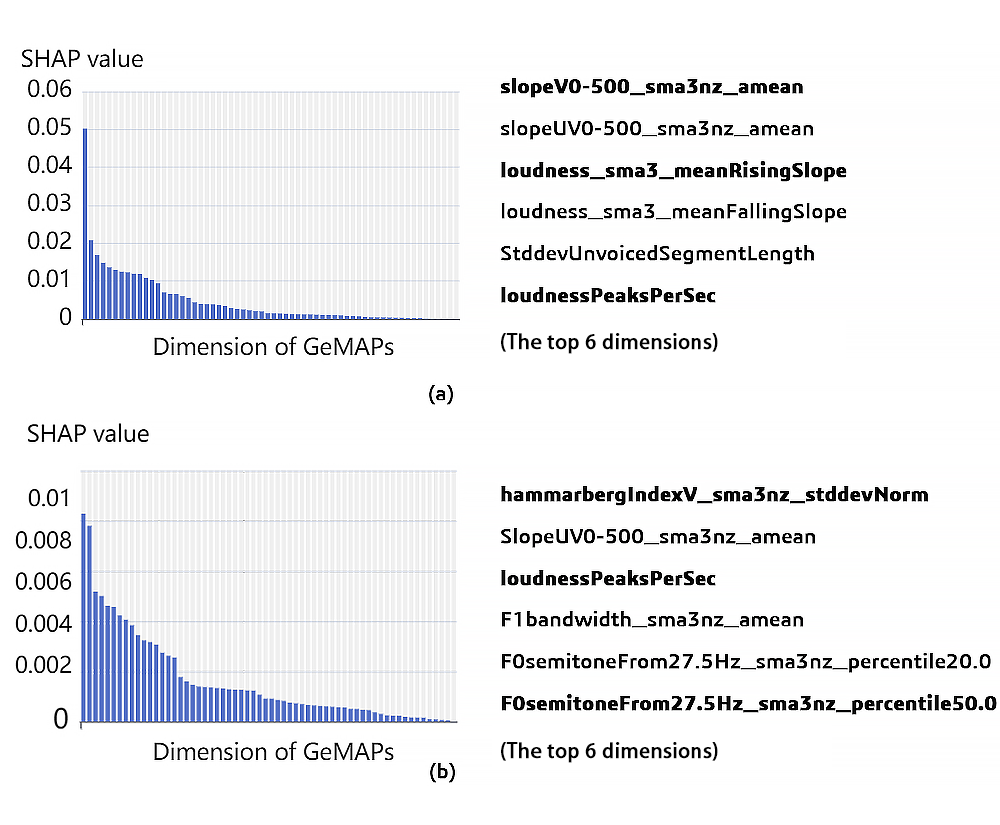}}
	\caption{(a) and (b) are the results of SHAP analysis ranked by descending SHAP values on open corpus and extracted human speech audios, respectively. The top six dimensions are shown on the right, 
and those that overlap with prominent factors in vocals of dogs are bolded.
}
	\label{fig:humanspeech}
\end{figure}

In order to better compare the vocals with human language, we conduct a SHAP analysis on 
human speech datasets: open public language corpus CommonVoice and the host speech 
extracted from the same videos as the dog vocals. 
To find out the difference between these two languages, we use xgboost as well to classify 
human speech into English or Japanese and then compute Shapley values, 
with results presented in \figref{fig:humanspeech}.

In the human speech analysis, the difference between English and Japanese concentrates on the slope, loudness, and F0semitone. From the above, we can conclude that the difference between vocals coming from different host language environments is mainly related to frequency. In the meantime, from our analysis of prominent acoustic factors, the vocalsvocals  of dogs and 
voices of humans share several common prominent factors, suggesting that the 
host human language have a correlation with the vocals of dogs.
\begin{table*}[th]
	\small
	\centering
	\begin{tabular}{l|c|c|c|c}
		\toprule
		\multirow{2}*{Dimension} &
		\multicolumn{2}{c|}{With Host Speech} & \multicolumn{2}{c}{With Random Speech}\\
		\cline{2-5} 
		& Pearson Coefficient & \textit{p}-value & Pearson Coefficient &  \textit{p}-value \\ \hline
		\textbf{loudness\_sma3\_amean}& 0.537& 	2.73e-83& 	0.025 & 0.405  \\ 
		\textbf{F0semitoneFrom27.5Hz\_sma3nz\_percentile50.0} & 0.105& 	4.76e-4	& -0.012 &	0.691\\ 
		\textbf{loudness\_sma3\_meanRisingSlope} & 0.159& 	1.08e-7& 	0.043& 	0.155 \\
		logRelF0-H1-A3\_sma3nz\_stddevNorm & 0.002 & 0.941	& 4.25e-4 & 0.989\\
		\textbf{loudnessPeaksPerSec} & 0.124 & 3.75e-5 & 1.20e-3 & 0.968\\
		\textbf{F0semitoneFrom27.5Hz\_sma3nz\_percentile80.0} & 0.162 & 6.40e-8 & -0.011	& 0.714\\
		hammarbergIndexV\_sma3nz\_stddevNorm & 9.13e-3 & 0.762	& 6.47e-3 & 0.830\\
		\textbf{slopeV0-500\_sma3nz\_amean}& 0.580	& 7.37e-100 & 	0.029& 	0.334\\
		\textbf{loudness\_sma3\_percentile80.0} & 0.436 & 3.03e-52 & 	0.030 & 	0.322\\
		slopeV500-1500\_sma3nz\_stddevNorm & 0.017 & 0.573& -4.40e-3  & 0.884\\
		\bottomrule
	
	\end{tabular}
	\caption{The correlation analysis is on two groups of data. In the first group, the analysis is on the correlation between dog vocals and their host speech, in the second group the correlation is between dog vocals and random speech. The dimensions of high correlation with p-value lower than 0.05 are bold.}
	\label{table:prominentpearson}
\end{table*}

Furthermore, to ascertain the correlation between vocals and speech more 
directly besides the overlap of their prominent factors, we calculate the 
Pearson's correlation between the two from the same videos 
in the ten prominent dimensions selected (\tabref{table:prominentpearson}). 

\subsubsection{Language Speed Comparison}
\label{sec:prominentfactor1}
Dog vocals under English and Japanese host language environments differ not only in these 
acoustic dimensions above, but a more intuitive dimension: \textit{speed}.

A formal measure of the speed of speech is the number of syllables in 
a given time duration. For speech from the same video as the dog voices, 
we apply an automatic speech 
recognition (ASR) model in Whisper~\cite{radford2022robust}, one of the state-of-the-art ASR models 
which smoothly transcribes the extracted speech into either English or Japanese texts. 
At the same time, CommonVoice already provides corresponding texts for their audio.
The segmentation of text into sequences of syllables is done by some subfunction of 
\textit{eSpeak}~\cite{duddington2012espeak}, which is a compact open source software speech synthesizer for 
English and other languages.
However, as there is no concept of ``syllable'' nor alphabet for Shiba 
Inu barks, 
we adopt another similar way for syllablizing dog clips. The dog clips are 
divided into sequences of syllable-like units according to an oscillator-based 
algorithm~\cite{rasanen2018pre}.
\begin{figure}[th]
	\centering
	\scalebox{0.20}{\includegraphics{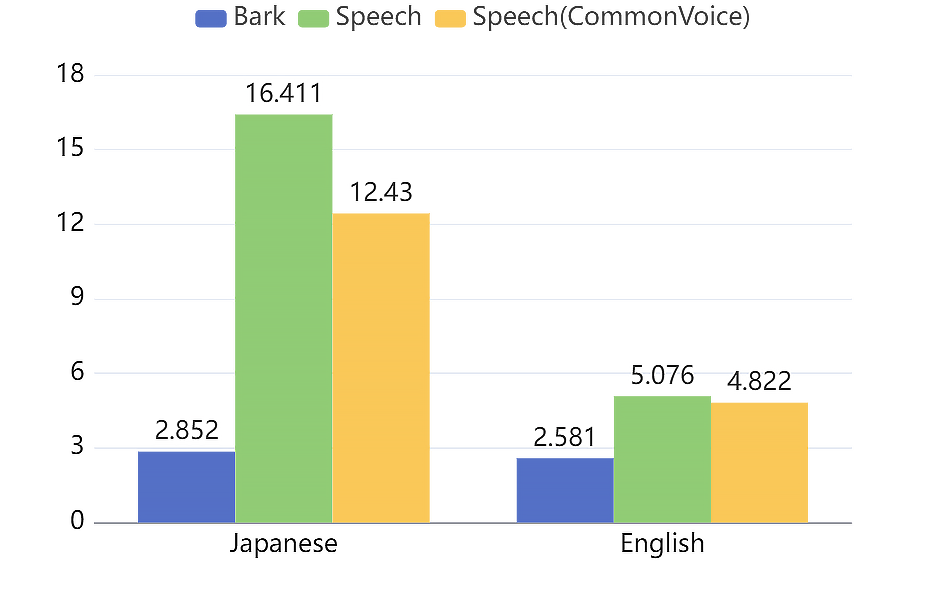}}
	\caption{The speed comparison for barks, host speech from same video and speech from CommonVoice.}
	\label{fig:speedcomparison}
\end{figure}

The number of syllables or syllable-like units per second is shown 
in \figref{fig:speedcomparison}.  We make the following discoveries: 
First, dog vocals from Japanese environment and Japanese human speech clips 
are generally faster than dog vocals from English environment and English speech clips. 
This suggests that when it comes to speed, there is a correlation between dog
vocals and human speech. 
Second, it appears that human speech in EJShibaVoice dataset is faster than
that of CommonVoice.  
The difference in speed can be attributed to the 
recording style of CommonVoice, which tends to be slower than typical speech.

\begin{figure}[th]
	\centering
\begin{subfigure}[t]{0.49\columnwidth}
        \centering
        \includegraphics[width=\textwidth]{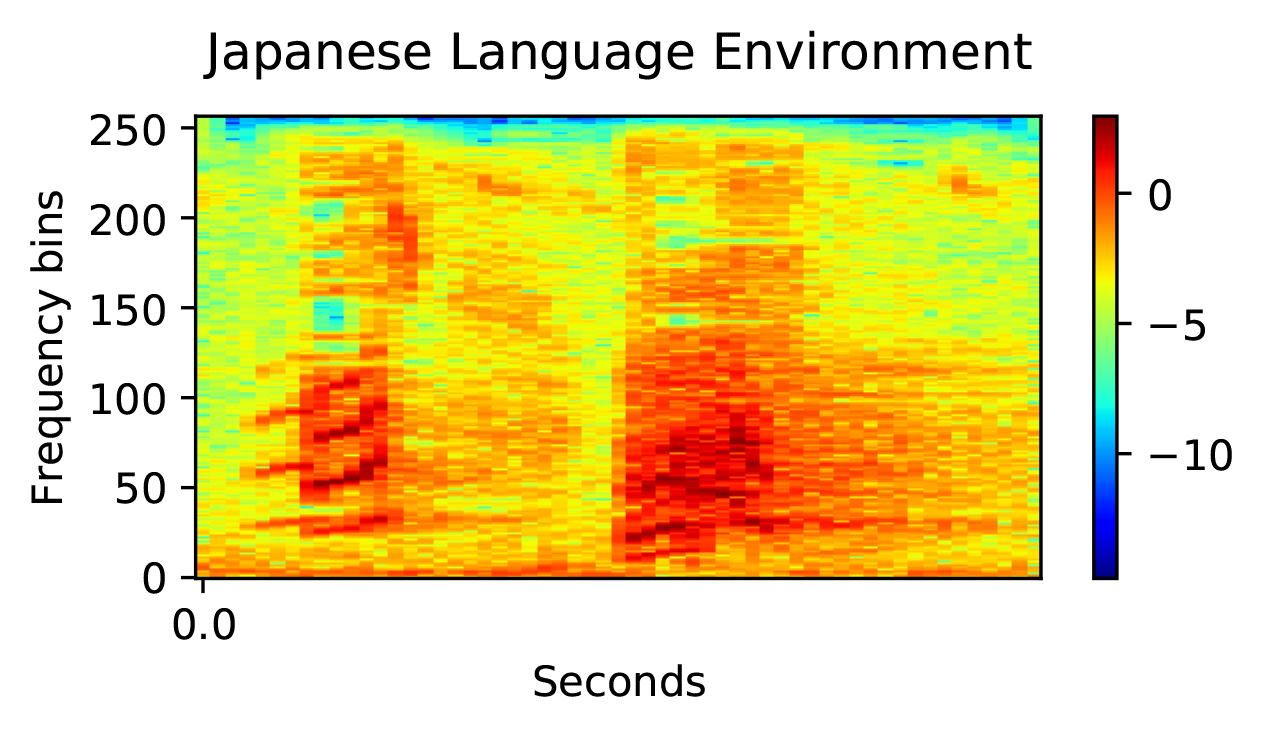}
        \caption{A Bark under Japanese Env.}
        \label{fig:subja}
\end{subfigure}
\begin{subfigure}[t]{0.49\columnwidth}
        \centering
        \includegraphics[width=\textwidth]{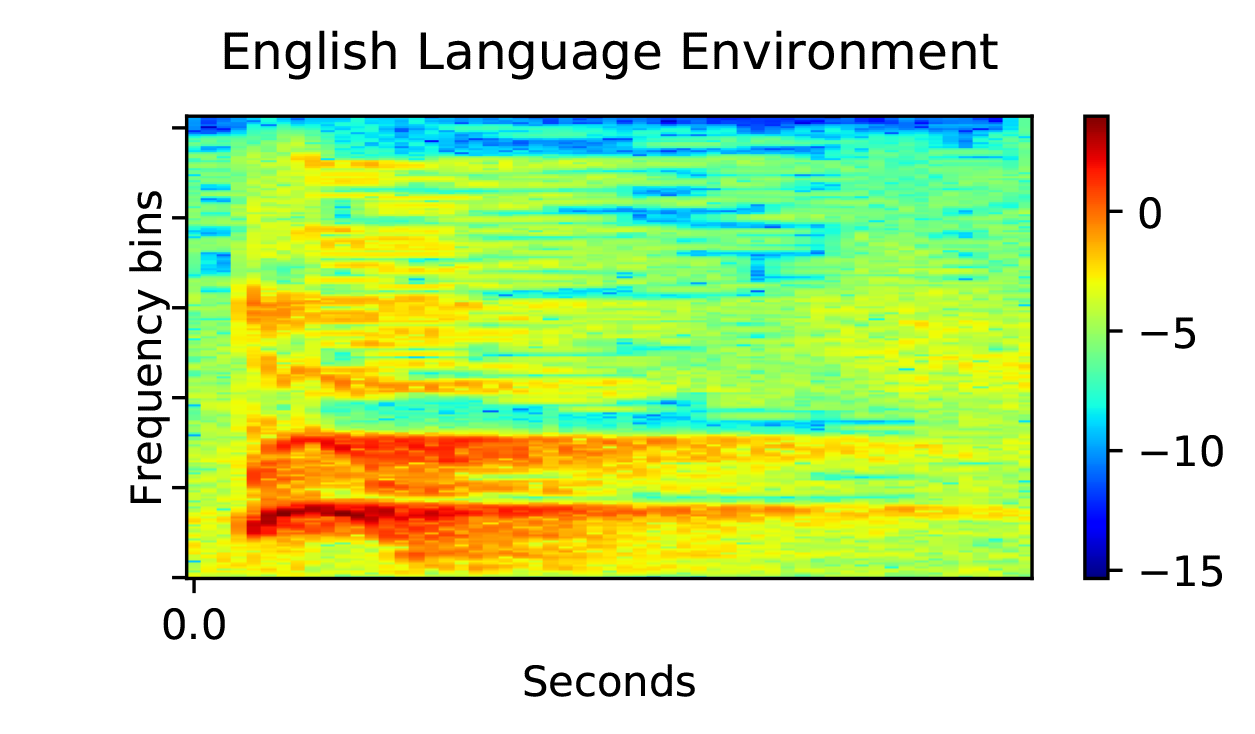}
        \caption{A Bark under English Env.}
        \label{fig:suben}
\end{subfigure}
	\caption{The spectrogram of two audio samples which are from different language environmentsand similar situations.}
	\label{Fig.sample}
\end{figure}

Here we illustrate the spectrograms of two typical dog vocal samples from the 
two language environments in~\figref{Fig.sample}. 
It is clear that in the same time length, the vocal under Japanese language environment has more syllable-like units than that under English language environment while the frequency of those vocals 
in the English environment is closer to the low-frequency region.

\subsection{Summary of Findings}

Through all the above analysis, we can find that \textbf{the key accoustic features that distinguish Shiba Inus from two language environments are frequency and speed, which correlate well with human speech.}



Our findings corroborate with previous literature~\cite{doi:10.1126/sciadv.aaw2594, graham2014fundamental} that the syllable speed of Japanese surpasses that of English and many other languages. At the same time, the frequency of Japanese is relatively higher than that of English. All these findings show that the vocals under English and Japanese language environments are different and in two dimensions of frequency and speed. To explain in detail, English dog vocals at a lower frequency than Japanese dogs, but Japanese dog vocals faster than English dogs. The same phenomenon can be observed with human speakers of these two languages. 





\subsection{Human Evaluation on Acoustic Features}
Alongside acoustic analysis, we assess cross-linguistic dog vocals via surveys. We pick 20 vocal pairs with distinct language settings from \secref{sec:method}. Each of the 20 questionnaires, given to 30 participants, makes 600 in total. Pair sequence is randomized. When spotting differences, participants gauge from four dimensions below
: (1)urgency, which represents the speed (2)pitch (3)duty ratio, which is the ratio of the duration of the voiced segments to that of the unvoiced segments (4)loudness. 

Among four above dimensions, 74.01\%, 79.09\%, 67.23\% and 63.56\% of the questionnaires which says that there are differences between the two clips report dimension difference respectively. This shows that from a human perspective the differences are most revealed by pitch, which is highly frequency related. This is in line with our acoustic analysis.



\subsection{Restrictions}
Compared to previous studies, we adopt a first-of-its-kind method of collecting data from the  Internet. Some may argue that former studies have a more controllable environment to eliminate confounding factors. However, it is very unlikely to cultivate exactly the same growing environment as subjectivity always exists. On the contrary, with the large amount of data from various families, we can remedy the confounding factors in a statistical way. Our pipeline is hence beneficial as we can scale up the data in the future and further validate our argument.

\section{Conclusion}
\label{sec:conclusion}
In this paper, we define a new problem of discovering the linguistic influence of 
the host on the sounds of their pet dogs. Experiments have shown that there is a 
significant difference in audio frequency and speed between the voice of dogs 
in the Japanese language environment and the English environment. 
Specifically, English dogs bark at a lower pitch than Japanese dogs, while Japanese dogs bark
faster than English ones. The phenomena can be observed in humans as well.

We sourced our data from YouTube, which, though noisy, ensures the quantity and variety and has 
rarely been covered in previous studies. The fact that we removed quite a lot of data from
the raw videos is due to the rigorous pipeline we developed for quality purposes. 
Eventually, our EJShibaVoice dataset, which contains a large number of Shiba Inu sound clips 
and their corresponding host speech under various scenes will facilitate future research in 
this field.   

Future direction can be a larger dataset for a more general investigation, i.e. more breeds and 
more sound clips, as well as more various language environments in addition to Japanese 
and English in this study. Note that the pipeline we proposed is language-agnostic and hence 
can be applied to other studies with similar purposes. 





\bibliography{sample-base}









\end{document}